\documentclass[aps,preprint,floats,epsf,epsfig,nofootinbib,letter]{revtex4}

\usepackage{graphicx}
\usepackage{dcolumn}
\usepackage{bm}

\begin{document}
\def\be{\begin{eqnarray}}
\def\en{\end{eqnarray}}
\def\non{\nonumber}
\def\la{\langle}
\def\ra{\rangle}
\def\nc{N_c^{\rm eff}}
\def\vp{\varepsilon}
\def\A{{\cal A}}
\def\B{{\cal B}}
\def\c{{\cal C}}
\def\d{{\cal D}}
\def\e{{\cal E}}
\def\p{{\cal P}}
\def\t{{\cal T}}
\def\N{{\cal N}}
\def\up{\uparrow}
\def\dw{\downarrow}
\def\vma{{_{V-A}}}
\def\vpa{{_{V+A}}}
\def\smp{{_{S-P}}}
\def\spp{{_{S+P}}}
\def\J{{J/\psi}}
\def\ol{\overline}
\def\ov{\overline}
\def\Lqcd{{\Lambda_{\rm QCD}}}
\def\pr{{\sl Phys. Rev.}~}
\def\prl{{\sl Phys. Rev. Lett.}~}
\def\pl{{\sl Phys. Lett.}~}
\def\np{{\sl Nucl. Phys.}~}
\def\zp{{\sl Z. Phys.}~}
\def\lsim{ {\ \lower-1.2pt\vbox{\hbox{\rlap{$<$}\lower5pt\vbox{\hbox{$\sim$}
}}}\ } }
\def\gsim{ {\ \lower-1.2pt\vbox{\hbox{\rlap{$>$}\lower5pt\vbox{\hbox{$\sim$}
}}}\ } }
\def\CP{{\it CP}~}
\newcommand{\acp}{\ensuremath{A_{CP}}}


\vskip 1.0 cm

\centerline{\large\bf On Charmless  $B\to K_h\eta^{(')}$ Decays }\centerline{\large\bf with $K_h=K,K^*,K_0^*(1430),K_2^*(1430)$}
\bigskip
\medskip
\centerline{\bf Hai-Yang Cheng,$^{1,2}$ Chun-Khiang Chua$^3$}
\bigskip
\centerline{$^1$ Institute of Physics, Academia Sinica}
\centerline{Taipei, Taiwan 115, Republic of China}
\medskip
\centerline{$^2$ Physics Department, Brookhaven National Laboratory} \centerline{Upton, New York 11973}
\medskip
\centerline{$^3$ Department of Physics, Chung Yuan Christian University}
\centerline{Chung-Li, Taiwan 320, Republic of China}
\bigskip

\medskip
\centerline{\bf Abstract}
\bigskip
\small

We study the charmless decays $B\to K_h\eta$ and $B\to K_h\eta'$  within the framework of QCD factorization (QCDF) for $K_h= K,K^*,K_0^*(1430)$ and naive factorization for $K_h=K_2^*(1430)$. There are three distinct types of penguin contributions: (i) $b\to sq\bar q\to s\eta_{q}$, (ii) $b\to ss\bar s\to s\eta_s$, and (iii) $b\to s q\bar q\to q \bar K_h$, where $\eta_q=(u\bar u+d\bar d)/\sqrt{2}$ and $\eta_s=s\bar s$. $B\to K^{(*)}\eta^{(')}$ decays are dominated by type-II and type-III penguin contributions. The interference, constructive for $K\eta'$ and $K^*\eta$ and destructive for $K\eta$ and $K^*\eta'$, between type-II and type-III diagrams explains the pattern of $\Gamma(B\to K\eta')\gg\Gamma(B\to K\eta)$ and $\Gamma(B\to K^*\eta')\ll\Gamma(B\to K^*\eta)$. Within QCDF, the observed large rate of the $K\eta'$ mode can be naturally explained without invoking flavor-singlet contributions or something exotic. The decay pattern for $B\to K_0^*(1430)\eta^{(')}$ decays   depends on whether the scalar meson $K_0^*(1430)$ is an excited state of $\kappa$ or a lowest-lying $P$-wave $q\bar q$ state.  Hence, the experimental measurements of $B\to K_0^*(1430)\eta^{(')}$ can be used to explore the quark structure of $K_0^*(1430)$. If $K_0^*(1430)$ is a low-lying $q\bar q$ bound state, we find that $K_0^*\eta$ has a rate slightly larger than $K_0^*\eta'$ owing to the fact that the $\eta$-$\eta'$ mixing angle in the $\eta_q,\eta_s$ flavor basis is less than $45^\circ$, in agreement with experiment.
Type-III penguin diagram does not contribute to $B\to K_2^*\eta^{(')}$ under the factorization hypothesis and type-II diagram dominates. The ratio $\Gamma(B\to K_2^*\eta')/\Gamma(B\to K_2^*\eta)$ is expected to be of order 2.5  as a consequence of (i) $|f_{\eta'}^s|>|f_\eta^s|$ and (ii) a destructive (constructive) interference between type-I and type-II penguin diagrams for $K^*_2\eta$ ($K_2^*\eta'$). However, the predicted rates of $B\to K_2^*\eta^{(')}$ in naive factorization are too small by one order of magnitude and this issue remains to be resolved. There are two $K^{(*)}\eta^{(')}$ modes in which direct \CP asymmetries have been measured with significance around $4\sigma$\,: $\acp(K^{-}\eta)=-0.37\pm0.09$ and $\acp(\bar K^{*0}\eta)=0.19\pm0.05$. In QCDF, power corrections from penguin annihilation which are needed to resolve \CP puzzles in $K^-\pi^+$ and $\pi^+\pi^-$ modes will flip $\acp(K^{-}\eta)$ into a wrong sign. We show that soft corrections to the color-suppressed tree amplitude $a_2$ in conjunction with the the charm content of the $\eta$ will finally lead to $\acp(K^-\eta)=-0.15^{+0.19}_{-0.28}$. Likewise, this power correction is needed to improve the prediction for $\acp(\bar K^{*0}\eta)$.

%
%

\begin{figure}[t]
\begin{center}
\includegraphics[width=0.80\textwidth]{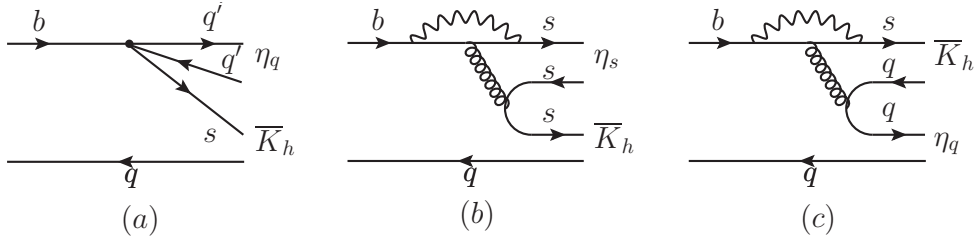}
\vspace{0.0cm}
\caption{Three different penguin contributions to $\ov B\to \ov K_h\eta^{(')}$ with $K_h$ denoting $K, K^*, K_0^*(1430)$ and $K_2^*(1430)$. Fig. 1(a) is induced by the penguin operators $O_{3,5,7,9}$.
} \label{fig:Keta} \end{center}
\end{figure}
\section{Introduction}
Recently BaBar has measured charmless $B$ decays with final states containing $\eta$ or $\eta'$ \cite{BaBar:Keta}. Comparing the first measurements of $B\to K_0^*(1430)\eta'$ and $B\to K_2^*(1430)\eta'$ by BaBar with previous results of $B\to K_0^*(1430)\eta$ and $B\to K_2^*(1430)\eta$ (see Table \ref{tab:data}) clearly indicates that $\B(B\to K_0^*(1430)\eta')<\B(B\to K_0^*(1430)\eta)$ and $\B(B\to K_2^*(1430)\eta')>\B(B\to K_2^*(1430)\eta)$. It is well known that $\B(B\to K\eta')\gg\B(B\to K\eta)$ and $\B(B\to K^*\eta')\ll\B(B\to K^*\eta)$. The last two patterns can be understood as the interference between the dominant penguin amplitudes.

\begin{table}[t]
\caption{Experimental branching fractions (in units of $10^{-6}$) of $B\to K_h\eta^{(')}$ with $K_h=K, K^*, K_0^*(1430)$ and $K_2^*(1430)$ taken from \cite{HFAG,BaBar:Keta}.
  \label{tab:data}}
\vspace{6pt}
\begin{ruledtabular}
\begin{tabular}{l c c c c c c c c}
 & $K\eta$ & $K\eta'$ & $K^*\eta$ & $K^*\eta'$ & $K_0^*\eta$ & $K_0^*\eta'$ & $K_2^*\eta$ & $K_2^*\eta'$  \\
\hline
$B^+$ & $2.36\pm0.27$ & $71.1\pm2.6$ & $19.3\pm1.6$ & $4.9^{+2.1}_{-1.9}$ & $15.8\pm3.1$ & $5.2\pm2.1$ & $9.1\pm3.0$ &  $28.0^{+5.3}_{-5.0}$ \\
$B^0$ & $1.12^{+0.30}_{-0.28}$ & $66.1\pm3.1$ & $15.9\pm1.0$ & $3.8\pm1.2$ & $9.6\pm1.9$ & $6.3\pm1.6$ & $9.6\pm2.1$ & $13.7^{+3.2}_{-3.1}$ \\
\end{tabular}
\end{ruledtabular}
\end{table}

For the $\eta$ and $\eta'$ particles,
it is more convenient to consider the flavor states   $q\bar q\equiv (u\bar u+d\bar
d)/\sqrt{2}$, $s\bar s$ and $c\bar c$ labeled by the $\eta_q$, $\eta_s$ and $\eta_{c}^0$, respectively. Neglecting the small mixing with $\eta_c^0$, we write
\begin{equation}\label{eq:qsmixing}
   \left( \begin{array}{c}
    |\eta\rangle \\ |\eta'\rangle
   \end{array} \right)
   = \left( \begin{array}{ccc}
    \cos\phi & -\sin\phi \\
    \sin\phi & \cos\phi
   \end{array} \right)
   \left( \begin{array}{c}
    |\eta_q\rangle \\ |\eta_s\rangle
   \end{array} \right) \;,
\end{equation}
where $\phi=(39.3\pm1.0)^\circ$ \cite{FKS} is the $\eta-\eta'$ mixing angle in the $\eta_q$ and $\eta_s$ flavor basis. Three different penguin contributions are depicted in Fig. \ref{fig:Keta}: (i) $b\to sq\bar q\to s\eta_{q}$, (ii) $b\to ss\bar s\to s\eta_s$, and (iii) $b\to s q\bar q\to q \bar K_h$, corresponding to Figs. 1(a), 1(b) and 1(c), respectively. For $B\to K^{(*)}\eta^{(')}$ decays, the dominant penguin amplitudes arise from Figs. 1(b) and 1(c) governed by the parameters $\alpha_4(K_h\eta_s)$ and $\alpha_4(\eta_q K_h)$, respectively. Their expressions in terms of the effective Wilson coefficients $a_4$ and $a_6$ are summarized in Table \ref{tab:alpha4}.

It is clear that the interference between the $B\to K\eta_q$ amplitude induced by the $b\to sq\bar q$ penguin and the $B\to K\eta_s$ amplitude induced by $b\to ss\bar s$  is constructive for $B\to K\eta'$ and destructive for $B\to K\eta$. This explains the large rate of the former and the suppression of the latter \cite{Lipkin}. For $B\to K^*\eta^{(')}$ decays, it is the other way around. The sign difference between $\alpha_4(\eta_q K^*)$ and $\alpha_4(K^*\eta_s)$ explains why $\Gamma(B\to K^*\eta)\gg \Gamma(B\to K^*\eta')$, recalling that $a_4$ and $a_6$ are negative and the magnitude of the latter is larger than the former and that the chiral factor $r_\chi$ to be defined below is of order unity for light mesons.

The decay pattern for $B\to K_0^*(1430)\eta^{(')}$ decays depends on whether $K_0^*(1430)$ is an excited state of $\kappa$ (or $K_0^*(800)$) or a low lying $P$-wave $q\bar q$ state. Hence, the experimental measurements of $B\to K_0^*(1430)\eta^{(')}$ can be used to explore the quark structure of the scalar meson $K_0^*(1430)$.
A detailed study in this work shows that $\Gamma(B\to K_0^*\eta)\ll \Gamma(B\to K^*_0\eta')$ in the first scenario for $K_0^*(1430)$ and $\Gamma(B\to K_0^*\eta)> \Gamma(B\to K^*_0\eta')$ in the latter scenario.
As for $B\to K_2^*(1430)\eta^{(')}$ decays,
Fig. 1(c) does not make contribution  owing to the vanishing decay constant of $K_2^*$. Since the interference between Figs. 1(a) and 1(b) is constructive for $K_2^*\eta'$ and destructive for $K_2^*\eta$ and since the decay constant $f_{\eta'}^s$ is larger than $f_\eta^s$, one will expect a larger rate for $K_2^*\eta'$ than $K_2^*\eta$.

\begin{table}[t]
\caption{The parameters $\alpha_4(K_h\eta_s)$ and $\alpha_4(\eta_q K_h)$  with $K_h=K, K^*, K_0^*(1430)$ and $K_2^*(1430)$.
  \label{tab:alpha4}}
\vspace{6pt}
\begin{ruledtabular}
\begin{tabular}{l c c c c}
& $K$ & $K^*$ & $K_0^*(1430)$ & $K_2^*(1430)$  \\
\hline
$\alpha_4(K_h\eta_s)$ & $a_4+r_\chi^{\eta_s}a_6$ & $a_4-r_\chi^{\eta_s}a_6$ & $a_4-r_\chi^{\eta_s}a_6$ & $a_4-r_\chi^{\eta_s}a_6$ \\
$\alpha_4(\eta_q K_h)$ & $a_4+r_\chi^{K}a_6$ & $a_4+r_\chi^{K^*}a_6$ & $a_4-r_\chi^{K_0^*}a_6$ & $-$ \\
\end{tabular}
\end{ruledtabular}
\end{table}

Recently we have studied the decays $B\to (K,K^*)\eta^{(')}$  within the framework of QCD factorization (QCDF) \cite{CC:Bud,CC:BCP}. Here we shall present updated results with some discussions. Then in the rest of this work we will focus on $B\to (K_0^*(1430),K^*_2(1430))\eta^{(')}$ decays and study their decay pattern.

The layout of the present paper is as follows. In Sec. II we recapitulate the framework of QCD factorization. The we proceed to study $B\to (K,K^*)(\eta,\eta')$ decays in Sec. III and $B\to K_0^*(1430)(\eta,\eta')$ decays in Sec. IV. Since the QCDF approach for the $K_2^*(1430)\eta^{(')}$ modes has not been developed, we reply on naive factorization to study the tensor meson production in Sec. V. Sec. VI comes to our conclusions. An appendix is devoted to the decay constants and matrix elements of the $\eta$ and $\eta'$ mesons.

\section{QCD factorization}

Within the framework of QCDF \cite{BBNS}, the effective
Hamiltonian matrix elements are written in the form
\begin{equation}\label{fac}
   \langle M_1M_2 |{\cal H}_{\rm eff}|\overline B\rangle
  \! =\! \frac{G_F}{\sqrt2}\sum_{p=u,c} \! \lambda_p^{(q)}\,
\!   \langle M_1M_2 |{\cal T_A}^{p}\!+\!{\cal
T_B}^{p}|\overline B\rangle \,,
\end{equation}
where $\lambda_p^{(q)}=V_{pb}V_{pq}^*$ with $q=s,d$, ${\cal
T_A}$ describes contributions from naive factorization, vertex
corrections, penguin contractions and spectator scattering expressed
in terms of the flavor operators $a_i^{p}$, while ${\cal T_B}$
contains annihilation topology amplitudes characterized by  the
annihilation operators $b_i^{p}$. The explicit expressions of ${\cal T_A}$ and ${\cal T_B}$ can be found in \cite{BN,BBNS}.
In practice, it is more convenient to express the decay amplitudes in terms of the flavor operators $\alpha_i^{p}$ and the annihilation operators  $\beta_i^p$. Their relations to the coefficients $a_i^{p}$ and $b_i^p$ will be specified below.

The expressions of $\bar B\to \bar K_h\eta^{(')}$ decay amplitudes  for $K_h=K,K^*,K_0^*(1430)$ and $K_2^*(1430)$ are given by \cite{BN}
\be \label{eq:KetaAmp}
\sqrt{2}A_{B^-\to K_h^-\eta^{(')}} &=& X^{(\bar B\bar K_h,\eta^{(')}_q)}\left[\delta_{pu}(\alpha_2+2\beta_{S2})+2\alpha_3^p+{1\over 2}\alpha^p_{3,EW}+2\beta^p_{S3}+2\beta^p_{S3,EW}\right] \non \\
&+& \sqrt{2}X^{(\bar B\bar K_h,\eta^{(')}_s)}\Big[ \delta_{pu}(\beta_2+2\beta_{S2})+\alpha_3^p+\alpha_4^p-{1\over 2}\alpha^p_{3,EW}-{1\over 2}\alpha^p_{4,EW}+\beta^p_{3}+\beta^p_{3,EW} \non \\
&+&\beta_{S3}^p+\beta^p_{S3,EW}\Big]+\sqrt{2}X^{(\bar B\bar K_h,\eta^{(')}_c)}\left[\delta_{pc}\alpha_2+\alpha_3^p\right] \non \\
&+& X^{(\bar B\eta^{(')}_q,\bar K_h)}\left[\delta_{pu}(\alpha_1+\beta_2)+\alpha_4^p+\alpha^p_{4,EW}+\beta_3^p+\beta^p_{3,EW}\right], \\
\sqrt{2}A_{\bar B^0\to \bar K_h^0\eta^{(')}} &=& X^{(\bar B\bar K_h,\eta^{(')}_q)}\left[\delta_{pu}\alpha_2+2\alpha_3^p+{1\over 2}\alpha^p_{3,EW}+2\beta^p_{S3}-\beta^p_{S3,EW}\right] \non \\
&+& \sqrt{2}X^{(\bar B\bar K_h,\eta^{(')}_s)}\Big[ \alpha_3^p+\alpha_4^p-{1\over 2}\alpha^p_{3,EW}-{1\over 2}\alpha^p_{4,EW}+\beta^p_{3}-{1\over 2}\beta^p_{3,EW}
+\beta_{S3}^p-{1\over 2}\beta^p_{S3,EW}\Big] \non \\
&+& \sqrt{2}X^{(\bar B\bar K_h,\eta^{(')}_c)}\left[\delta_{pc}\alpha_2+\alpha_3^p\right]
+X^{(\bar B\eta^{(')}_q,\bar K_h)}\left[\alpha_4^p-{1\over 2}\alpha^p_{4,EW}+\beta_3^p-{1\over 2}\beta^p_{3,EW}\right]. \non
\en
The order of the arguments of $\alpha_i^p (M_1 M_2)$ and $\beta_i^p(M_1
M_2)$, which are not shown explicitly here,  is consistent with the order of the arguments of the factorizable matrix elements
$X^{(\overline B M_1, M_2)}$ given by
\be
\label{eq:X}
X^{(\bar B P_1,P_2)} &\equiv& \la P_2|J^{\mu}|0\ra\la P_1|J'_{\mu}|\ov B\ra=if_{P_2}(m_{B}^2-m^2_{P_1}) F_0^{ B P_1}(m_{P_2}^2),  \non \\
X^{(\bar BP,V)} &\equiv & \la V| J^{\mu}|0\ra\la
P|J'_{\mu}|\ov B \ra=2f_V\,m_Bp_c F_1^{ B
P}(m_{V}^2),   \non \\
X^{( \bar BV,P)} &\equiv &
\la P | J^{\mu}|0\ra\la V|J'_{\mu}|\ov B
\ra=2f_P\,m_Bp_cA_0^{B V}(m_{P}^2),   \\
X^{(\bar BP,S)} &\equiv & \la S| J^{\mu}|0\ra\la
P|J'_{\mu}|\ov B \ra=f_S\,(m_B^2-m_P^2)F_0^{ B
P}(m_{S}^2),   \non \\
X^{( \bar BS,P)} &\equiv &
\la P | J^{\mu}|0\ra\la S|J'_{\mu}|\ov B
\ra=-f_P\,(m_B^2-m_S^2)F_0^{B S}(m_{P}^2), \non \\
X^{(\bar BT,P)} &\equiv & \la P | J^{\mu}|0\ra\la T|J'_{\mu}|\ov B
\ra=-if_P [k(m_P^2)+(m_B^2-m_T^2)b_+(m_P^2)+m_P^2 b_-(m_P^2)] \vp^*_{\mu\nu}p_B^\mu p_B^\nu, \non \\
X^{(\bar BP,T)} &\equiv & \la T | J^{\mu}|0\ra\la P|J'_{\mu}|\ov B
\ra=0, \non
\en
where $f_P,f_V,f_S$ are the decay constants of pseudoscalar, vector and scalar mesons, respectively, and $k,b_+,b_-$ are $B$ to tensor meson transition form factors defined in Eq. (\ref{eq:BTff}) below.

The flavor operators $\alpha_i^{p}$ and the annihilation operators  $\beta_i^p$ are related to the coefficients $a_i^{p}$ and $b_i^p$ by
\begin{eqnarray}\label{eq:alphai}
   \alpha_1(M_1 M_2) &=& a_1(M_1 M_2) \,, \nonumber\\
   \alpha_2(M_1 M_2) &=& a_2(M_1 M_2) \,, \nonumber\\
   \alpha_3^{p}(M_1 M_2) &=& \left\{
    \begin{array}{cl}
     a_3^{p}(M_1 M_2) - a_5^{p}(M_1 M_2)
      & \quad \mbox{for~} M_1 M_2=PP, \, VP, \, SP,\, TP \\
     a_3^{p}(M_1 M_2) + a_5^{p}(M_1 M_2)
      & \quad \mbox{for~} M_1 M_2=PV,\, PS ,
    \end{array}\right. \nonumber\\
   \alpha_4^{p}(M_1 M_2) &=& \left\{
    \begin{array}{cl}
     a_4^{p}(M_1 M_2) + r_{\chi}^{M_2}\,a_6^{p}(M_1 M_2)
      & \quad \mbox{for~} M_1 M_2=PP, \, PV , \\
     a_4^{p}(M_1 M_2) - r_{\chi}^{M_2}\,a_6^{p}(M_1 M_2)
      & \quad \mbox{for~} M_1 M_2=VP, \, SP, \, PS,\, TP
    \end{array}\right.\\
   \alpha_{3,\rm EW}^{p}(M_1 M_2) &=& \left\{
    \begin{array}{cl}
     a_9^{p}(M_1 M_2) - a_7^{p}(M_1 M_2)
      & \quad \mbox{for~} M_1 M_2=PP, \, VP ,\, SP, \,TP\\
     a_9^{p}(M_1 M_2) + a_7^{p}(M_1 M_2)
      & \quad \mbox{for~} M_1 M_2=PV, \, PS,
    \end{array}\right. \nonumber\\
   \alpha_{4,\rm EW}^{p}(M_1 M_2) &=& \left\{
    \begin{array}{cl}
     a_{10}^{p}(M_1 M_2) + r_{\chi}^{M_2}\,a_8^{p}(M_1 M_2)
      & \quad \mbox{for~} M_1 M_2=PP, \, PV , \\
     a_{10}^{p}(M_1 M_2) - r_{\chi}^{M_2}\,a_8^{p}(M_1 M_2)
      & \quad \mbox{for~} M_1 M_2=VP\,, SP,\, PS,\, TP
     \end{array}\right. \nonumber
\end{eqnarray}
and
\be \label{eq:beta}
 \beta_i^p (M_1 M_2) =\frac{i f_B f_{M_1}
f_{M_2}}{X^{(\overline B M_1,M_2)}}b_i^p,
\en
where the chiral factors $r_\chi$'s are given by
\be \label{eq:rchi}
&& r_\chi^P(\mu)={2m_P^2\over m_b(\mu)(m_2+m_1)(\mu)},  \quad
   r_\chi^{\eta_s}= {h_P^s\over f_P^s m_b(\mu) m_s(\mu)},\non \\
&& r_\chi^V(\mu) = \frac{2m_V}{m_b(\mu)}\,\frac{f_V^\perp(\mu)}{f_V}, \qquad\qquad r_\chi^{K_0^*}(\mu)={2m_{K_0^*}^2\over m_b(\mu)(m_s-m_q)(\mu)},
\en
with the parameters $f_P^s$ and $h_P^s$ being defined in the Appendix.

The flavor operators $a_i^{p}$ are basically the Wilson coefficients
in conjunction with short-distance nonfactorizable corrections such
as vertex corrections and hard spectator interactions. In general,
they have the expressions \cite{BBNS,BN}
 \be \label{eq:ai}
  a_i^{p}(M_1M_2) =
 \left(c_i+{c_{i\pm1}\over N_c}\right)N_i(M_2)
  + {c_{i\pm1}\over N_c}\,{C_F\alpha_s\over
 4\pi}\Big[V_i(M_2)+{4\pi^2\over N_c}H_i(M_1M_2)\Big]+P_i^{p}(M_2),
 \en
where $i=1,\cdots,10$,  the upper (lower) signs apply when $i$ is
odd (even), $c_i$ are the Wilson coefficients,
$C_F=(N_c^2-1)/(2N_c)$ with $N_c=3$, $M_2$ is the emitted meson
and $M_1$ shares the same spectator quark with the $B$ meson. The
quantities $V_i(M_2)$ account for vertex corrections,
$H_i(M_1M_2)$ for hard spectator interactions with a hard gluon
exchange between the emitted meson and the spectator quark of the
$B$ meson and $P_i(M_2)$ for penguin contractions. The expression
of the quantities $N_i(M_2)$ reads
 \be
 N_i(M_2)=\cases{0, & $i=6,8$ and $M_2=V$, \cr
                 1, & {\rm else}. \cr}
 \en

In Eq. (\ref{eq:KetaAmp}), possible flavor-singlet penguin annihilation contributions are denoted by $\beta_S$'s which will not be considered in this work.

Power corrections in QCDF always involve troublesome endpoint divergences. For
example, the annihilation amplitude has endpoint divergences even at twist-2 level and the hard spectator scattering diagram at twist-3 order is power
suppressed and posses soft and collinear divergences arising from the soft
spectator quark. Since the treatment of endpoint divergences is model dependent, subleading power corrections generally can be studied only in a
phenomenological way. We shall follow \cite{BBNS} to model the endpoint divergence $X\equiv\int^1_0 dx/(1-x)$ in the annihilation and hard spectator
scattering diagrams as
 \be \label{eq:XA}
 X_A=\ln\left({m_B\over \Lambda_h}\right)(1+\rho_A e^{i\phi_A}), \qquad
 X_H=\ln\left({m_B\over \Lambda_h}\right)(1+\rho_H e^{i\phi_H}),
 \en
with $\Lambda_h$ being a typical scale of order
500 MeV, and  $\rho_{A,H}$, $\phi_{A,H}$ being the unknown real parameters.

As pointed out in \cite{CC:BCP}, while the discrepancies between experiment  and theory in the heavy quark limit for the  rates of penguin-dominated two-body decays of $B$ mesons and direct \CP asymmetries of $\bar B^0\to K^-\pi^+$, $B^-\to K^-\rho^0$ and $\bar B^0\to \pi^+\pi^-$ are resolved by introducing power corrections coming from penguin annihilation, the signs of  direct {\it CP}-violating effects in  $B^-\to K^-\pi^0, B^-\to K^-\eta$ and $\bar B^0\to\pi^0\pi^0$ are flipped to the wrong ones when confronted with experiment. These new $B$-{\it CP} puzzles in QCDF can be resolved by the subleading power corrections to the color-suppressed tree amplitudes due to spectator interactions and/or final-state interactions \cite{Chua} that not only reproduce correct signs for  aforementioned \CP asymmetries but also accommodate the observed $\bar B_d\to \pi^0\pi^0$ and $\rho^0\pi^0$ rates simultaneously.
Following \cite{CC:BCP}, power corrections to the color-suppressed topology are parametrized as \be \label{eq:a2}
a_2 \to a_2(1+\rho_C e^{i\phi_C}),
\en
with the unknown parameters $\rho_C$ and $\phi_C$ to be inferred from experiment.

For $S$-wave mesons, input parameters  such as decay constants, form factors, quark masses, Wolfenstein parameters, light-cone distribution amplitudes, power correction parameters can be found in \cite{CC:Bud}. Input parameters for parity-even mesons such as $K_0^*(1430)$ and $K_2^*(1430)$ will be specified later. For the renormalization scale
of the decay amplitude, we choose $\mu=m_b(m_b)=4.2$ GeV.

\section{$B\to (K,K^*)(\eta,\eta')$ decays}

 \begin{table}[tbp!]
 \caption{Branching fractions (top; in units of $10^{-6}$) and direct {\it CP} asymmetries (bottom; in units of \%) of  $B\to (K,K^*)(\eta,\eta')$ decays obtained in various approaches. The pQCD results are taken from \cite{XiaoKeta} for $B\to K\eta^{(')}$ with partial NLO corrections and from \cite{ChenKeta} for $B\to K^*\eta^{(')}$.
 There are two solution sets with SCET predictions for decays involving $\eta$ and/or $\eta'$ \cite{Zupan,SCETVP}. The theoretical errors correspond to the uncertainties due to the variation of (i) Gegenbauer moments, decay constants, quark masses, form factors, the $\lambda_B$ parameter for the $B$ meson wave function, and (ii) $\rho_{A,H}$, $\phi_{A,H}$, respectively.
 } \label{tab:PPBr}
\begin{ruledtabular}
 \begin{tabular}{l  c c c c c}
 {Mode}
   &  QCDF (this work) & pQCD &  SCET  & Expt. \cite{HFAG,BaBar:Keta} \\  \hline
  $B^-\to K^-\eta$                                               &
                                                                 $2.2^{+1.7+1.1}_{-1.0-0.9}$ &
                                                                 $3.2^{+1.2+2.7+1.1}_{-0.9-1.2-1.0}$ & $2.7\pm4.8\pm0.4\pm0.3$ & $2.36\pm0.27$  \\
                                                                 &&&$2.3\pm4.5\pm0.4\pm0.3$ &\\
  $B^-\to K^-\eta'$
                                                                 &   $74.5^{+57.9+25.6}_{-25.3-19.0}$
                                                                 &
                                                                 $51.0^{+13.5+11.2+4.2}_{-~8.2-~\,6.2-3.5}$ & $69.5\pm27.0\pm4.4\pm7.7$ & $71.1\pm2.6$  \\
                                                                 &&&$69.3\pm26.0\pm7.1\pm6.3$
                                                                 &\\
  $\bar B^0\to \bar K^0\eta$
                                                                 & $1.5^{+1.4+0.9}_{-0.8-0.7}$ &
                                                                 $2.1^{+0.8+2.3+1.0}_{-0.6-1.0-0.9}$ & $2.4\pm4.4\pm0.2\pm0.3$& $1.12^{+0.30}_{-0.28}$  \\
                                                                 &&&$2.3\pm4.4\pm0.2\pm0.5$
                                                                 &\\
  $\bar B^0\to \bar K^0\eta'$                                    &
                                                                 $70.9^{+54.1+24.2}_{-23.8-18.0}$ &
                                                                 $50.3^{+11.8+11.1+4.5}_{-~8.2-~6.2-2.7}$ &$63.2\pm24.7\pm4.2\pm8.1$ & $66.1\pm3.1$ & \\
                                                                 &&&$62.2\pm23.7\pm5.5\pm7.2$
                                                                 &\\
  $B^-\to K^{*-}\eta $                           & $15.8^{+8.2+9.6}_{-4.2-7.3}$
                                    &                                    $22.13^{+0.26}_{-0.27}$
                                    & $17.9_{-5.4-2.9}^{+5.5+3.5}$
                                    & $19.3\pm1.6$ &\\
                                    &&& $18.6_{-4.8-2.2}^{+4.5+2.5}$   &\\
  $B^-\to K^{*-}\eta' $                    & $1.6^{+2.1+3.7}_{-0.3-1.6}$
                                    & $6.38\pm0.26$
                                    & $4.5_{-3.9-0.8}^{+6.6+0.9}$
                                    & $4.8^{+1.8}_{-1.6}$ \footnotemark[1] &\\
                                    && & $4.8_{-3.7-0.6}^{+5.3+0.8}$ & \\
  $\bar B^0\to \bar K^{*0}\eta$                   & $15.7^{+7.7+9.6}_{-4.0-7.3}$
                                    & $22.31^{+0.28}_{-0.29}$
                                    & $16.6_{-5.0-2.7}^{+5.1+3.2}$
                                    & $15.9\pm1.0$ & \\
                                    &&& $16.5_{-4.3-2.0}^{+4.1+2.3}$ &\\
  $\bar B^0\to \bar K^{*0}\eta'$                 & $1.5^{+1.8+3.5}_{-0.3-1.6}$
                                    & $3.35^{+0.29}_{-0.27}$
                                    & $4.1_{-3.6-0.7}^{+6.2+0.9}$                                 & $3.1^{+1.0}_{-0.9}$ \footnotemark[2]  & \\
                                    && & $4.0_{-3.4-0.6}^{+4.7+0.7}$ &\\
                                    \hline
  $B^-\to K^-\eta$                                               &
                                                                 $-14.5^{+10.3+15.5}_{-26.0-10.7}$ & $-11.7^{+6.8+3.9+2.9}_{-9.6-4.2-5.6}$ & $33\pm30\pm7\pm3$ & $-37\pm9$  \\
                                                                 &&&$-33\pm39\pm10\pm4$
                                                                 &\\
  $B^-\to K^-\eta'$                                              &
                                                                 $0.45^{+0.69+1.20}_{-0.55-0.98}$ & $-6.2^{+1.2+1.3+1.3}_{-1.1-1.0-1.0}$ &$-10\pm6\pm7\pm5$ & $1.3^{+1.6}_{-1.7}$ \\
                                                                 &&&$0.7\pm0.5\pm0.2\pm0.9$
                                                                 &\\
  $\bar B^0\to \bar K^0\eta$                                     &
                                                                 $-23.6^{+~9.8+12.6}_{-26.2-12.5}$ & $-12.7^{+4.1+3.2+3.2}_{-4.1-1.5-6.7}$ & $21\pm20\pm4\pm3$ & \\
                                                                 &&&$-18\pm22\pm6\pm4$
                                                                 &\\
  $\bar B^0\to \bar K^0\eta'$                                    &
                                                                 $3.0^{+0.6+0.8}_{-0.5-0.8}$ & $2.3^{+0.5+0.3+0.2}_{-0.4-0.6-0.1}$ &$11\pm6\pm12\pm2$ & $5\pm5$ \\ &&&$-27\pm7\pm8\pm5$ &\\
  $B^-\to K^{*-}\eta $                                 & $-10.1^{+3.9+6.5}_{-3.7-7.8}$
                                    & $-24.57^{+0.72}_{-0.27}$
                                    & $-2.6_{-5.5-0.3}^{+5.4+0.3}$
                                    & $2\pm6$ & \\
                                    && & $-1.9_{-3.6-0.1}^{+3.4+0.1}$ &\\
  $B^-\to K^{*-}\eta' $                       & $69.7^{+~6.5+27.9}_{-38.6-49.5}$
                                    & $4.60^{+1.16}_{-1.32}$
                                    & $2.7_{-19.5-0.3}^{+27.4+0.4}$

                                    & $-26\pm27$ & \\
                                    &&  & $2.6_{-32.9-0.2}^{+26.7+0.2}$ &\\
  $\bar B^0\to \bar K^{*0}\eta$                       & $3.4^{+0.4+2.7}_{-0.4-2.4}$
                                    & $0.57\pm0.011$
                                    & $-1.1_{-2.4-0.1}^{+2.3+0.1}$
                                    & $19\pm5$ & \\
                                    && & $-0.7_{-1.3-0.0}^{+1.2+0.1}$ &\\
  $\bar B^0\to \bar K^{*0}\eta'$                 & $8.8^{+~8.8+30.8}_{-10.7-24.1}$
                                    & $-1.30\pm0.08$
                                    & $9.6_{-11.0-1.2}^{+~8.9+1.3}$
                                    &  $2\pm23$  &  \\
                                    && & $9.9_{-4.3-0.9}^{+6.2+0.9}$ &\\
\end{tabular}
\footnotetext[1]{This is from the BaBar data \cite{BaBar:Keta}. Belle obtained an upper limit  $2.9\times 10^{-6}$ \cite{Belle:Ksteta'}.}
\footnotetext[2]{This is from the BaBar data \cite{BaBar:Keta}. Belle obtained an upper limit  $2.6\times 10^{-6}$ \cite{Belle:Ksteta'}.}
\end{ruledtabular}
 \end{table}

\subsection{Branching fractions}

Details of the calculations in the framework of QCDF for all $B\to PP,VP$ decays can be found in \cite{CC:Bud,CC:BCP}. The updated results for the branching fractions and direct \CP asymmetries in $B\to K^{(*)}\eta^{(')}$ decays are exhibited in Table \ref{tab:PPBr} after correcting some minor errors in the previous computer codes.

Numerically, Beneke and Neubert already obtained $\B(B^-\to K^-\eta')\sim {\cal O}(50\times 10^{-6})$ in QCDF using the default values $\rho_A=\rho_H=0$ \cite{BN}. Here we found similar results $57\times 10^{-6}$ ($53\times 10^{-6}$) with (without) the contributions from the ``charm content" of the $\eta'$. In the presence of penguin annihilation, we obtain $\B(B^-\to K^-\eta')\sim 75\times 10^{-6}$ ($67\times 10^{-6}$)
with (without) the ``charm content" contributions.
Therefore, the observed large $B\to K\eta'$ rates are naturally explained in QCDF without invoking, for example, significant flavor-singlet contributions or an enhanced hadronic matrix element $\la 0|\bar s\gamma_5 s|\eta'\ra$. Data on $B\to K\eta$ modes are also well accounted for by QCDF.

The values of the parameters $\alpha_4(K^{(*)}\eta_{q,s})$ and $\alpha_4(\eta_qK^{(*)})$ are given by
\be
\alpha_4(K\eta_s)=-0.098-0.013i, &\qquad & \alpha_4(\eta_q K)=-0.094-0.013i, \non \\
\alpha_4(K^*\eta_s)=0.035-0.0009i, &\qquad & \alpha_4(\eta_q K^*)=-0.036-0.0085i.
\en
The magnitude of $\alpha_4(\eta_q K^*)$ is smaller than $\alpha_4(\eta_q K)$ owing to the smallness of $r_\chi^{K^*}\sim 0.36$ compared to $r_\chi^K\sim 1.45$ at the scale $\mu=4.2$ GeV, while the smallness of $\alpha_4(K^*\eta_s)$ relative to $\alpha_4(K\eta_s)$ is due to the destructive interference in the former. The sign difference between $\alpha_4(\eta_q K^*)$ and $\alpha_4(K^*\eta_s)$ explains why $\Gamma(B\to K^*\eta)\gg \Gamma(B\to K^*\eta')$.  Although the rates of $K^*\eta'$ and $K\eta$ are comparable, $\B(B\to K^*\eta)$ is much smaller than $\B(B\to K\eta')$.

The QCDF prediction for the branching fraction of $B\to K^*\eta'$, of order $1.5\times 10^{-6}$, \footnote{The predictions $\B(B^-\to K^{*-}\eta')=2.2\times 10^{-6}$ and $\B(\bar B^0\to \bar K^{*0}\eta')=1.9\times 10^{-6}$ obtained by Beneke and Neubert \cite{BN} in the so-called ``S4" scenario  in which power corrections to penguin annihilation are taken into account are consistent with ours.}
is smaller than the predictions of pQCD and soft-collinear effective theory (SCET), but it is consistent with experiment within errors. The experimental values quoted in Table \ref{tab:PPBr} are  the BaBar measurements \cite{BaBar:Keta}. Belle obtained only the upper bounds: $\B(B^-\to K^{*-}\eta')<2.9\times 10^{-6}$ and $\B(\bar B^0\to \bar K^{*0}\eta')<2.6\times 10^{-6}$ \cite{Belle:Ksteta'}. Therefore, although our central values are smaller than BaBar, they are consistent with Belle. It is very important to measure them to discriminate between various model predictions.

\subsection{$\CP$ asymmetries}

There are two modes in which direct \CP asymmetries have been measured with significance around $4\sigma$\,: $\acp(K^{-}\eta)=-0.37\pm0.09$ and $\acp(\bar K^{*0}\eta)=0.19\pm0.05$. It is crucial  to understand them. Since the two penguin diagrams Figs. 1(b) and 1(c) contribute
destructively to $B\to K\eta$ due to the opposite sign of $f_\eta^q$ and $f_\eta^s$, the penguin amplitude is comparable in magnitude to the tree amplitude induced from $b\to us\bar u$, contrary to the decay $B\to K\eta'$ which is dominated by large penguin amplitudes. Consequently, a sizable direct \CP asymmetry is expected in $B^-\to K^-\eta$ but not in $K^-\eta'$ \cite{BSS}.

In the absence of any power corrections, it appears that the QCDF prediction $\acp(K^-\eta)=-0.233^{+0.164}_{-0.193}$  obtained in the leading $1/m_b$ expansion already agrees well with the data. \footnote{Beneke and Neubert \cite{BN} obtained $\acp(K^-\eta)=-0.189^{+0.290}_{-0.300}$ in their default predictions.}
However, this agreement is just an accident. Recall that when power corrections are turned off, the predicted \CP asymmetries for the penguin-dominated modes $K^-\pi^+$, $K^{*-}\pi^+$, $K^-\rho^+$, $K^-\rho^0$, and tree-dominated modes $\pi^+\pi^-$, $\rho^\pm\pi^\mp$ and $\rho^-\pi^+$ are wrong in signs when confronted with experiment \cite{CC:Bud,CC:BCP}. That is why it is important to consider the effects of power corrections step by step. The QCDF results in the heavy quark limit should  not be considered as the final QCDF predictions to be compared with experiment.
It turns out that the aforementioned wrong signs can be flipped into the right direction by the power corrections from penguin annihilation.

However, a scrutiny of the QCDF predictions reveals more puzzles with respect to direct \CP violation. While the signs of \CP asymmetries in $K^-\pi^+,K^-\rho^0$ modes etc., are flipped to the right ones in the presence of power corrections from penguin annihilation,
the signs of $\acp$ in $B^-\to K^-\pi^0,~K^-\eta,~\pi^-\eta$ and $\bar B^0\to\pi^0\pi^0,~\bar K^{*0}\eta$ will also get reversed in such a way that they disagree with experiment \cite{CC:Bud,CC:BCP}. Specifically,
$\acp(K^-\eta)$ is found to be of order $0.11$ in the presence of penguin annihilation and hence it has a wrong sign.  These \CP puzzles can be resolved by having soft corrections to the color-suppressed tree coefficient $a_2$ so that $a_2$ is large and complex.
When $\rho_C$ and $\phi_C$ are turned on (see Eq. (\ref{eq:a2}) with $\rho_C=1.3$ and $\phi_C=-70^\circ$ \cite{CC:Bud}), $\acp(K^-\eta)$ will be reduced to $-0.02$
if there is no  intrinsic charm content of the $\eta$. Although the decay constant $f_\eta^c\approx -2$ MeV is much smaller than $f_\eta^{q,s}$ [see Eq. (\ref{eq:fetaqs})], its effect is CKM enhanced by $V_{cb}V_{cs}^*/(V_{ub}V_{us}^*)$. Therefore, the charm content of the $\eta$ may have a significant impact on \CP violation. Indeed, when $f_\eta^c$ is turned on, $\acp(K^-\eta)$ finally reaches the level of $-15\%$ with a sign in agreement with experiment. Hence, \CP asymmetry in $B^-\to K^-\eta$ is the place where the charm content of the $\eta$ plays a role. \footnote{One of us (C.K.C.) has studied residual final-state interaction effects in charmless $B$ decays \cite{Chua}. In this approach, $\acp(K^-\eta)$ of order $-0.27$ can be induced through the decay $B^-\to K^-\eta'$ followed by $K^-\eta'\to K^-\eta$ rescattering via penguin diagrams. This rescattering mimics the effect of the charm content in the $\eta$.}
The pQCD prediction is similar to QCDF. For comments on the pQCD calculation of $\acp(K^-\eta)$, the reader is referred to \cite{CC:Bud}. Note that while both QCDF and pQCD lead to a correct sign for $\acp(K^-\eta)$, the predicted magnitude still falls short of the measurement $-0.37\pm0.09$.

As for \CP asymmetry in $\bar B^0\to\bar K^{*0}\eta$, we have $\acp(\bar K^{*0}\eta)=(3.1^{+2.0}_{-1.9})\%$ in the heavy quark limit. It is modified to $(0.12^{+2.10}_{-1.60})\%$ by penguin annihilation. Soft corrections to the color-suppressed tree amplitude is needed to improve the prediction and finally
we obtain $\acp(\bar K^{*0}\eta)=(3.4^{+2.8}_{-2.4})\%$ (Table \ref{tab:PPBr}). Unlike the $K^0\eta$ mode, the charm content of the $\eta$ here does not play an essential role for \CP violation in $K^{*0}\eta$.
We see from Table \ref{tab:PPBr} that QCDF is in better agreement with experiment than pQCD and SCET, though it is still smaller than the data.

\section{$B\to K_0^*(1430)\eta^{(')}$ decays in QCDF}
The hadronic charmless $B$ decays into a scalar meson and a pseudoscalar meson have been studied in QCDF in \cite{CCY:SP}. For scalar mesons above 1 GeV we have explored two possible scenarios in the QCD sum rule method, depending on
whether the light scalars $\kappa,~a_0(980)$ and $f_0(980)$ are
treated as the lowest lying $q\bar q$ states or four-quark
particles: (i) In scenario 1, we treat
$\kappa, a_0(980), f_0(980)$ as the lowest lying states, and
$K_0^*(1430), a_0(1450),f_0(1500)$ as the corresponding first
excited states, respectively,  and (ii) we assume in scenario 2 that $K_0^*(1430),
a_0(1450), f_0(1500)$ are the lowest lying resonances and the
corresponding first excited states lie between $(2.0\sim
2.3)$~GeV. Scenario 2 corresponds to the case that light scalar
mesons are four-quark bound states, while all scalar mesons are
made of two quarks in scenario 1.  We found that scenario 2 is preferable. Indeed, lattice calculations have confirmed that $a_0(1450)$ and $K_0^*(1430)$ are lowest-lying $P$-wave $q\bar q$ mesons \cite{Mathur}, and suggested that $\sigma$ and $\kappa$ are $S$-wave tetraquark mesonia \cite{Prelovsek,Mathur}.

Decay constants of scalar mesons are defined as
 \be \label{eq:Sdecayc}
 \la S(p)|\bar q_2\gamma_\mu q_1|0\ra=f_S p_\mu,
 \qquad \la S|\bar q_2q_1|0\ra=m_S\bar f_S.
 \en
the vector decay constant $f_S$ and the
scale-dependent scalar decay constant $\bar f_S$ are related by
equations of motion
 \be \label{eq:EOM}
 \mu_Sf_S=\bar f_S, \qquad\quad{\rm with}~~\mu_S={m_S\over
 m_2(\mu)-m_1(\mu)},
 \en
where $m_{2}$ and $m_{1}$ are the running current quark masses and
$m_S$ is the scalar meson mass.
In general, the twist-2 light-cone distribution amplitude (LCDA) of the scalar meson $\Phi_S$
has the form
 \be \label{eq:twist2wf}
 \Phi_S(x,\mu)=f_S\,6x(1-x)\left[1+\mu_S\sum_{m=1}^\infty
 B_m(\mu)\,C_m^{3/2}(2x-1)\right],
 \en
where $B_m$ are Gegenbauer moments and $C_m^{3/2}$ are the
Gegenbauer polynomials. For twist-3 LCDAs, we use
 \be \label{eq:twist3wf}
 \Phi^s_S(x)=\bar f_S , \qquad \Phi^\sigma_S(x)=\bar f_S\,
 6x(1-x).
 \en
Since $\mu_S\equiv 1/B_0\gg 1$ and even
Gegenbauer coefficients are suppressed, it is clear that the LCDA
of the scalar meson is dominated by the odd Gegenabuer moments. In
contrast, the odd Gegenbauer moments vanish for the $\pi$ and
$\rho$ mesons. The Gegenbauer moments $B_1$, $B_3$ and the scalar decay constant $\bar f_{K_0^*}$  in scenarios 1
and 2  obtained using the QCD sum rule method \cite{CCY:SP} are listed in Table \ref{tab:momentscenario1}.

\begin{table}[tb]
\caption{Gegenbauer moments $B_1$, $B_3$ and the scalar decay constant $\bar f_{K_0^*}$ (in units of MeV) in scenario 1
and scenario 2  at the scales $\mu=1$ GeV and 2.1 GeV
(shown in parentheses) obtained using the QCD sum rule method
\cite{CCY:SP}. } \label{tab:momentscenario1}
\begin{ruledtabular}
\begin{tabular}{l c c c}
 & $B_1$ & $B_3$ & $\bar f_{K_0^*}$ \\ \hline
 scenario 1 &$0.58\pm 0.07~(0.39\pm 0.05)$ & $-1.20\pm 0.08~ (-0.70\pm  0.05)$ & $-300\pm30~(-370\pm35)$\\
 scenario 2 &
 $-0.57\pm 0.13~ (-0.39\pm 0.09)$ & $-0.42\pm 0.22~ (-0.25\pm 0.13)$ & $445\pm50~(550\pm60)$\\
\end{tabular}
\end{ruledtabular}
\end{table}

\begin{table}[tb]
\caption{Values of $\alpha_{2,3,4}$ at $\mu=4.2$ GeV in scenarios 1 and 2. } \label{tab:WC}
\begin{ruledtabular}
\begin{tabular}{l c c c c c c}
 & $\alpha_2(K_0^*\eta_{q,s})$ & $\alpha_3(K_0^*\eta_{q,s})$ & $\alpha_4^c(K_0^*\eta_{s})$ & $\alpha_4^c(K_0^*\eta_{q})$ & $\alpha_4^c(\eta_{q}K_0^*)$ \\ \hline
 scenario 1 & $-0.27-0.08i$ & $0.050+0.006i$ & $0.025+0.001i$ & $0.0001-0.0017i$ & $0.50+0.09i$ \\
 scenario 2 & $0.51-0.08i$ & $-0.047+0.006i$ & $0.043+0.009i$ & $0.018-0.002i$ & $0.45+0.02i$ \\
\end{tabular}
\end{ruledtabular}
\end{table}

Form factors for $B\to S$ transitions are defined by
\be \label{eq:FF}
\la S(p')|A_\mu|B(p)\ra &=& -i\Bigg[\left(P_\mu-{m_B^2-m_S^2\over
q^2}\,q_ \mu\right) F_1^{BS}(q^2)   +{m_B^2-m_S^2\over
q^2}q_\mu\,F_0^{BS}(q^2)\Bigg],
 \en
where $P_\mu=(p+p')_\mu$, $q_\mu=(p-p')_\mu$.  As shown in
\cite{CCH}, a factor of $(-i)$ is needed in $B\to S$ transition in
order to obtain positive $B\to S$ form factors. This also can be
checked from heavy quark symmetry \cite{CCH}. As a consequence, the factorizable amplitudes $X^{(BP,S)}$ and $X^{(BS,P)}$ defined in Eq. (\ref{eq:X}) are of opposite sign. In this work we shall use the form factors \footnote{In footnote 9 of \cite{CCY:SP},
we have emphasized that the decay constant and the form factor for the excited state are of opposite sign. Hence, we do not agree with \cite{XiaoKeta,Lu} on the sign of $F_0^{BK_0^*}$ in scenario 1.}
\be
F_0^{BK_0^*}(q^2)={0.21\over 1-0.59(q^2/m_B^2)+0.09(q^4/m_B^4)}
\en
in scenario 1
and
\be
F_0^{BK_0^*}(q^2)={0.26\over 1-0.44(q^2/m_B^2)+0.05(q^4/m_B^4)}
\en
in scenario 2.
They are obtained using the covariant light-front quark model \cite{CCH}.

For the convenience of ensuing discussions, we would write
\be
A_{B\to K_0^*\eta} &=& X^{(BK_0^*,\eta_q)} C_1+X^{(BK_0^*,\eta_s)}C_2+X^{(B\eta_q,K_0^*)}C_3, \non \\
A_{B\to K_0^*\eta'} &=& X^{(BK_0^*,\eta'_q)}C_1+X^{(BK_0^*,\eta'_s)}C_2+X^{(B\eta'_q,K_0^*)}C_3,
\en
where $C_1$, $C_2$ and $C_3$ terms correspond to Figs. 1(a), 1(b) and 1(c), respectively. The expressions of $C_i$'s in terms of the parameters $\alpha_i^p$ can be found in Eq. (\ref{eq:KetaAmp}).
As noticed in \cite{CCY:SP}, the pattern of $a_i$ or $\alpha_i$ in $B\to SP$ decays (see Table \ref{tab:WC}) can be quite different from that in $PP$ modes. For example, we have
\be
&& \alpha_2(K \eta_{q,s})=0.21-0.08i,
\qquad\qquad \alpha_3(K\eta_{q,s})=-0.010-0.0058i, \\
&& \alpha_4^c(K\eta_{q})=-0.073-0.011i, \quad \alpha_4^c(K\eta_{s})=-0.098-0.013i, \quad \alpha_4^c(\eta_q K)=-0.094-0.013i \non
\en
for $K\eta^{(')}$ modes. Since the chiral factor $r_\chi^{K^*_0}= 12.3$ at $\mu=4.2$ GeV is larger than $r_\chi^K=1.5$ by one order of magnitude
owing to the large mass of $K_0^*(1430)$, it follows that $\alpha_4^c(\eta_q K_0^*)$ is much greater than $\alpha_4^c(K_0^*\eta_s)$ and $\alpha_4^c(K_0^*\eta_q)$.
Likewise, the real part of $\alpha_{2,3}(K_0^*\eta_{q,s})$ is greater than that of $\alpha_{2,3}(K\eta_{q,s})$ because the hard spectator term $H(M_1 M_2)$
\begin{eqnarray}\label{eq:hardspec}
  H(M_1 M_2)= {if_B f_{M_1} f_{M_2} \over X^{(\overline{B} M_1,
  M_2)}}\,{m_B\over\lambda_B} \int^1_0 d x d y \,
 \Bigg( \frac{\Phi_{M_1}(x) \Phi_{M_2}(y)}{(1-x)(1-y)} + r_\chi^{M_1}
  \frac{\Phi_{m_1} (x) \Phi_{M_2}(y)}{(1-x) y}\Bigg),
 \hspace{0.5cm}
 \end{eqnarray}
with $\Phi_M$ and $\Phi_m$ being the twist-2 and twist-3 LCDAs of the meson $M$, respectively,
is greatly enhanced for $M_1=K_0^*(1430)$ due to the large chiral factor $r_\chi^{K_0^*}$.

Because of the small vector decay constant of $K_0^*(1430)$, $X^{(B\eta^{(')}_q,K_0^*)}$ is suppressed relative to
$X^{(BK_0^*,\eta^{(')}_q)}$ and $X^{(BK_0^*,\eta^{(')}_s)}$. However, $C_3$ gains a large enhancement from $\alpha_4^c(\eta_q K_0^*)$. As a result, the amplitude of Fig. 1(c) is comparable to that of Fig. 1(a). The decay pattern of $B\to K_0^*\eta^{(')}$ depends on whether the scalar meson $K_0^*(1430)$ is an excited state of $\kappa$ or a lowest-lying $P$-wave $q\bar q$ state. In scenario 1, we have (in units of GeV$^3$)
\be
&& X^{(BK_0^*,\eta_q)}=-0.60, \quad X^{(BK_0^*,\eta_s)}=0.61, \qquad X^{(B\eta_q,K_0^*)}=-0.15, \non \\
&& X^{(BK_0^*,\eta_q')}=-0.49, \quad X^{(BK_0^*,\eta_s')}=-0.75, \quad~ X^{(B\eta_q',K_0^*)}=-0.12\,.
\en
Consequently, Fig. 1(b) interferes destructively (constructively) with Figs. 1(a) and 1(c) for $K_0^*\eta$ ($K_0^*\eta'$). This leads to $\B(B\to K_0^*\eta')\gg \B(B\to K_0^*\eta)$ (see Table \ref{tab:SPBr}), which is in sharp disagreement with experiment.

The decay pattern in scenario 2 is quite different.
Because of the large magnitude of  $\alpha_3(K_0^*\eta_{q,s})$ and the large cancellation between $\alpha_3(K_0^*\eta_s)$ and $\alpha_4(K_0^*\eta_s)$ in $C_2$, $B\to K_0^*\eta^{(')}$ decays are dominated by the contributions from Figs. 1(a) and 1(c), contrary to the $B\to K^{(*)}\eta^{(')}$ decays which are governed by Figs. 1(b) and 1(c). Numerically, we obtain
\be
 C_1=-0.098+0.011i, &&\quad X^{(BK_0^*,\eta_q)}=-0.74, \quad X^{(B\eta_q, K_0^*)}=0.22, \non \\
 C_3=0.455+0.017i, && \quad X^{(BK_0^*,\eta'_q)}=-0.60, \quad X^{(B\eta'_q, K_0^*)}=0.18\,.
\en
Therefore, the penguin diagrams Figs. 1(a) and 1(c) contribute constructively to both $K_0^*\eta$ and $K^*_0\eta'$ with comparable magnitudes. Since $X^{(BK_0^*,\eta_q)}/X^{(BK_0^*,\eta'_q)}=X^{(B\eta_q, K_0^*)}/X^{(B\eta'_q, K_0^*)}=\cot\phi\approx 1.23$, it is clear that $A_{B\to K_0^*\eta}/A_{B\to K_0^*\eta'}\approx \cot\!\phi$ and hence $B\to K_0^*\eta$ should have a rate larger than $B\to K_0^*\eta'$ in scenario 2 as the mixing angle $\phi$ is less than $45^\circ$.

 \begin{table}[tbp!]
 \caption{Branching fractions (top; in units of $10^{-6}$) and direct {\it CP} asymmetries (bottom; in units of \%) of  $B\to K_0^*(1430)(\eta,\eta')$ decays obtained in QCDF (this work) and pQCD \cite{Xiao:K0steta} in scenario 1 (first entry) and scenario 2 (second entry).
 } \label{tab:SPBr}
\begin{ruledtabular}
 \begin{tabular}{l  c c c c c}
 {Mode}
   &  QCDF (this work) & pQCD & Expt. \cite{HFAG,BaBar:Keta} \\  \hline
  $B^-\to K_0^{*-}(1430)\eta$                                               &
                                                                 $0.5^{+1.7+11.9}_{-0.4-~0.5}$ &
                                                                 $11.8^{+5.3+0.3+1.1+2.5}_{-3.5-0.4-1.2-2.3}$ & $15.8\pm3.1$  \\
                                                                 & $13.3^{+6.2+48.6}_{-3.6-10.1}$ &$33.8^{+13.5+1.1+7.7+8.2}_{-~9.0-1.1-7.0-7.3}$ &\\
  $B^-\to K_0^{*-}(1430)\eta'$
                                                                 &   $27.8^{+20.8+14.1}_{-10.2-17.6}$
                                                                 &
                                                                 $21.6^{+1.6+3.1+4.0+4.5}_{-0.5-2.8-3.6-4.1}$ & $5.2\pm2.1$  \\
                                                                 &$9.6^{+12.9+10.0}_{-~6.1-~9.4}$ &$77.5^{+15.8+6.2+21.0+18.0}_{-10.8-5.8-16.5-16.1}$ & \\
  $\bar B^0\to \bar K_0^{*0}(1430)\eta$
                                                                 & $0.4^{+1.5+11.8}_{-0.3-~0.4}$ &
                                                                 $9.1^{+4.4+0.0+1.1+2.0}_{-2.8-0.1-1.1-1.8}$ & $9.6\pm1.9$  \\
                                                                 & $12.6^{+5.7+50.1}_{-3.4-~9.3}$
                                                                 &$28.4^{+11.6+1.4+6.4+6.9}_{-~7.8-1.4-5.9-6.2}$ & \\
  $\bar B^0\to \bar K_0^{*0}(1430)\eta'$                                    &
                                                                 $26.6^{+19.7+17.3}_{-~9.8-16.5}$ &
                                                                 $22.0^{+1.6+3.2+3.9+4.6}_{-0.5-3.6-3.0-4.2}$ & $6.3\pm1.6$ & \\
                                                                 & $8.7^{+11.7+7.5}_{-~5.5-3.1}$ &$74.2^{+15.0+6.4+20.5+17.2}_{-10.3-5.7-16.2-15.5}$\\
                                                                 \hline
  $B^-\to K_0^{*-}(1430)\eta$                                               &
                                                                 $8.3^{+21.1+11.7}_{-~5.4-11.7}$ & & $5\pm13$ \\
                                                                 & $1.7^{+0.2+2.4}_{-0.2-2.4}$ \\
  $B^-\to K_0^{*-}(1430)\eta'$                                              &
                                                                 $-0.3^{+0.2+0.4}_{-0.2-0.4}$ & & $6\pm20$ \\
                                                                 & $1.2^{+0.7+1.7}_{-0.6-1.7}$ \\
  $\bar B^0\to \bar K_0^{*0}(1430)\eta$                                     &
                                                                 $14.5^{+36.3+20.4}_{-10.8-20.4}$ & & $6\pm13$ \\
                                                                 &$2.2^{+0.2+3.1}_{-0.2-3.1}$ \\
  $\bar B^0\to \bar K_0^{*0}(1430)\eta'$                                    &
                                                                 $-0.2^{+0.1+0.3}_{-0.1-0.3}$ & & $-19\pm17$ \\
                                                                 &$1.1^{+0.5+1.6}_{-0.5-1.6}$
                                                                 \\
\end{tabular}
\end{ruledtabular}
 \end{table}

The QCDF results for branching fractions and \CP asymmetries of $B\to K_0^*(1430)\eta^{(')}$ obtained in two different scenarios are shown in Table \ref{tab:SPBr} where the central values are for $\rho_A=\rho_C=0$ and $\phi_A=\phi_C=0$ and the ranges $0\leq \rho_{A,H}\leq 0.5$ and $0\leq \phi_{A,H}\leq 2\pi$ have been considered for the estimate of uncertainties.
It is evident that scenario 2 is in better agreement with experiment than scenario 1. This implies that the scalar meson $K_0^*(1430)$ is a lowest lying rather than excited $q\bar q$ state.

Several remarks are in order. (i) If $X^{(BK_0^*,\eta_q)}$ and $X^{(B\eta_q,K_0^*)}$ are of the same sign, the interference between $C_1$ and $C_3$ terms will become destructive in scenario 2 and yield too small rates for both $K_0^*\eta$ and $K_0^*\eta'$.
(ii) A recent pQCD calculation \cite{XiaoKeta} indicates that $\B(B\to K_0^*\eta')/\B(B\to K_0^*\eta)\approx 2$ in scenario 1 and $\B(B\to K_0^*\eta')\approx 75\times 10^{-6}$ in scenario 2. Neither of them is consistent with experiment. (iii)
Different estimates of form factors at $q^2=0$ were obtained recently: $F_0^{BK_0^*}(0)\sim -0.34$ \cite{Lu}, $-0.44$ \cite{XiaoKeta} in scenario 1 and $F_0^{BK_0^*}(0)\sim 0.60$ \cite{Lu}, 0.76 \cite{XiaoKeta} in scenario 2. If these form factors are used in QCDF calculations, we will have $\B(B\to K_0^*\eta)\sim 22\times 10^{-6}\gg \B(B\to K_0^*\eta')\sim 1\times 10^{-6}$ in scenario 2 for $F_0^{BK_0^*}(0)=0.60$ as an example.
This implies that a small form factor for $B\to K_0^*(1430)$ transition is preferable.

\section{$B\to K_2^*(1430)\eta^{(')}$ in naive factorization}

The observed $J^P=2^+$ tensor mesons $f_2(1270)$, $f_2'(1525)$, $a_2(1320)$ and $K_2^*(1430)$ form an SU(3) $1\,^3P_2$ nonet.  The $q\bar q$ content for isodoublet and isovector tensor resonances is obvious.
The polarization tensor $\vp_{\mu\nu}$ of a $^3P_2$ tensor meson with $J^{PC}=2^{++}$ satisfies the relations
 \be
 \vp_{\mu\nu}=\vp_{\nu\mu} ~, \qquad \vp^{\mu}_{~\mu}=0 ~, \qquad
 p_\mu \vp^{\mu\nu}=p_\nu\vp^{\mu\nu}=0 ~,
 \en
where $p^\mu$ is the momentum of the tensor meson.  Therefore,
 \be
 \la 0|(V-A)_\mu|T(\vp,p)\ra = a\vp_{\mu\nu}p^\nu+b\,\vp^\nu_{~\nu} p_\mu=0 ~,
 \en
and hence the decay constant of the tensor meson vanishes identically; that is, the tensor meson cannot be produced from the $V-A$ current. This means that Fig. 1(c) does not contribute to $B\to K_2^*\eta^{(')}$ under the factorization hypothesis.

\begin{table}[t]
\caption{Parameters in the form factors of $B\to K_2^*(1430)$ transitions in the parametrization of Eq.~(\ref{eq:FFpara}), as obtained by fitting to the covariant light-front model \cite{CCH}. The form factor $k$ is dimensionless, while $k$, $b_+$ and $b_-$ are in units
of ${\rm GeV}^{-2}$. The numbers in parentheses are the form factors at $q^2=0$ obtained using the ISGW2 model \cite{ISGW2}. }
 \label{tab:FFBtoT}
\begin{ruledtabular}
\begin{tabular}{| l c  c c || l c  c c |}
~$F$~~~~~
    & $F(0)$~~~~~
    & $a$~~~~~
    & $b$~~~~~~
& ~ $F$~~~~~
    & $F(0)$~~~~~
    & $a$~~~~~
    & $b$~~~~~~
 \\
    \hline
$h^{BK_2^*}$
    & $0.008~(0.016)$
    &  2.17
    &  2.22
&$k^{BK_2^*}$
    & $0.015~(0.293)$
    & $-3.70$
    & $1.78$
    \\
$b_+^{BK_2^*}$
    & $-0.006~(-0.006)$
    &  1.96
    &  1.79 &
$b_-^{BK_2^*}$
    & $0.002~(0.0063)$
    & $0.38$
    & $0.92$
    \\
\end{tabular}
\end{ruledtabular}
\end{table}

The general expression for the $B\to T$ transition has the form \cite{ISGW}
 \be \label{eq:BTff}
 \la T(\vp,p_T)|(V-A)_\mu|B(p_B)\ra &=&
 ih(q^2)\epsilon_{\mu\nu\rho\sigma}\vp^{*\nu\alpha}p_{B\alpha}(p_B+p_T)^\rho
 (p_B-p_T)^\sigma+k(q^2)\vp^*_{\mu\nu}p_B^\nu  \non \\
 &+& b_+(q^2)\vp^*_{\alpha\beta}p_B^\alpha p_B^\beta(p_B+p_T)_\mu
 +b_-(q^2)\vp^*_{\alpha\beta}p_B^\alpha p_B^\beta(p_B-p_T)_\mu.
 \en
The form factors $h$, $k$, $b_+$ and $b_-$ have been calculated in the ISGW quark model \cite{ISGW} and its improved version, the ISGW2 model \cite{ISGW2}.  They have also been computed in the covariant light-front (CLF) quark model \cite{CCH} and are listed in Table~\ref{tab:FFBtoT} for form factors fitted to a 3-parameter
form
  \be \label{eq:FFpara}
 F(q^2)=\,{F(0)\over 1-a(q^2/m_{B}^2)+b(q^2/m_{B}^2)^2}\,.
 \en
Notice that when $q^2$ increases, $h(q^2)$ ,  $b_+(q^2)$ and $b_-(q^2)$ increases more rapidly in the CLF model than in the ISGW2 model and that the form factor $k$ in both models is quite different.

The decay amplitude of $B\to TP$ always has the generic expression
 \be
 A(B\to TP)
 =\vp^*_{\mu\nu}p_B^\mu p_B^\nu\,M(B\to TP) ~.
 \en
The decay rate is given by
 \be \label{eq:rateTP}
 \Gamma(B\to TP)=\,{p_c^5\over 12\pi m_T^2}
 \left({m_B\over m_T}\right)^2|M(B\to TP)|^2 ~,
 \en
where $p_c$ is the magnitude of the 3-momentum of either final-state meson in the rest frame of the $B$ meson. Since the QCDF approach for $B\to TP$ has not been developed, we will reply on naive factorization to make estimates; that is, we will not include vertex, hard spectator and penguin corrections in Eq. (\ref{eq:ai}) for the calculation of $a_i$.

The predictions of $B\to K_2^*(1430)\eta^{(')}$ in various models are shown in Table \ref{tab:BrTP}. We see that the predicted rates in naive factorization are too small by one order of magnitude. \footnote{Although the form factor $k(q^2)$ is very different in the CLF and ISGW2 models, the magnitude of $F(q^2)=k(q^2)+(m_B^2-m_T^2)b_+(q^2)+q^2b_-(q^2)=-0.15$ in the former and $0.14$ in the latter for $q^2$ in the range between $m_\eta^2$ and $m_{\eta'}^2$ is about the same. Consequently, the predicted rates of $B\to K_2^*\eta^{(')}$ are basically independent of the form-factor models.}
It was found in \cite{Kim} that $\B(B\to K_2^*\eta')/\B(B\to K_2^*\eta)\sim 45$, a prediction not borne out by experiment. This is ascribed to the fact that the matrix element
\be
\la \eta'|\bar s\gamma_5 s|0\ra=-i{m_{\eta'}^2\over 2m_s}f_{\eta'}^s,
\en
used in \cite{Kim} is erroneous as it does not have the correct chiral behavior in the chiral limit; the SU(3)-singlet $\eta_1$ acquires a mass of the QCD anomaly which does not vanish in the chiral limit. Applying the anomalous equation of motion Eq. (\ref{eq:eom}) and neglecting the masses of $u$ and $d$ quarks, one gets \cite{Kagan,Ali}
\be \label{eq:etame}
\la \eta^{(')}|\bar s\gamma_5 s|0\ra=-i{m_{\eta^{(')}}^2\over 2m_s}\left(f_{\eta^{(')}}^s-{1\over\sqrt{2}}f_{\eta^{(')}}^q\right).
\en
Since $f_{\eta'}^q/\sqrt{2}$ is about half of $f_{\eta'}^s$, this means that the $\eta'$ matrix element of the pseudoscalar density is reduced roughly by a factor of 2. On the contrary, the $\eta$ matrix element is enhanced owing to the opposite sign of $f_\eta^q$ and $f_\eta^s$.
A rigorous result without neglecting light quark masses is $\la \eta^{(')}|\bar s\gamma_5 s|0\ra=-ih_{\eta^{(')}}^s/ (2m_s)$ [cf. Eq. (\ref{eq:hetaqs})]. Since the magnitude of $h_\eta^s$ is numerically close to $h_{\eta'}^s$, see Eq. (\ref{eq:fetaqs}), it becomes clear that the rate of $K^*_2\eta$ is comparable to but slightly smaller than the $K_2^*\eta'$ one.
Our results agree with \cite{Munoz} as we use the same form factors derived from the CLF model. \footnote{The only difference between our work and \cite{Munoz} is that we use Eq. (\ref{eq:hetaqs}) for the $\eta^{(')}$ matrix elements of pseudoscalar densities, while the authors of \cite{Munoz} use Eq. (\ref{eq:etame}).}
The predictions of \cite{Verma} are too small as the authors only considered the tree diagram effects and neglect the important penguin contributions.

 \begin{table}[tbp!]
 \caption{Branching fractions (in units of $10^{-6}$) of  $B\to K_2^*(1430)(\eta,\eta')$ decays obtained in various approaches. The predictions of \cite{Kim} are for $1/N_c^{\rm eff}=0.3$\,.
 } \label{tab:BrTP}
\begin{ruledtabular}
 \begin{tabular}{l  c c c c c}
 {Mode}
   &  This work & KLO \cite{Kim} & MQ \cite{Munoz} & SV \cite{Verma} & Expt. \cite{HFAG,BaBar:Keta} \\  \hline
  $B^-\to K_2^{*-}(1430)\eta$                                               &
                                                                 $1.1$ &
                                                                 $0.031$ & $1.19$ & $0.01$
                                                                  & $9.1\pm3.0$
                                                                 \\
  $B^-\to K_2^{*-}(1430)\eta'$
                                                                 &   $2.7$
                                                                 &
                                                                 $1.4$ & $2.70$  & $0.007$ & $28.0^{+5.3}_{-5.0}$ \\
  $\bar B^0\to \bar K_2^{*0}(1430)\eta$
                                                                 & $1.0$ &
                                                                 $0.029$ & $1.09$ & $0.01$ & $9.6\pm2.1$ \\
  $\bar B^0\to \bar K_2^{*0}(1430)\eta'$                                    &
                                                                 $2.5$ &
                                                                 $1.3$ & $2.46$ & $0.006$ & $13.7^{+3.2}_{-3.1}$ \\
\end{tabular}
\end{ruledtabular}
 \end{table}

A slightly large rate of $K_2^*\eta'$ over $K_2^*\eta$ comes from two sources: (i) $|f_{\eta'}^s|>|f_\eta^s|$ and (ii) a destructive (constructive) interference between Figs. 1(a) and 1(b) for $K^*_2\eta$ ($K_2^*\eta'$). It is expected that $\Gamma(B\to K_2^*\eta')/\Gamma(B\to K_2^*\eta)\sim 2.5$\,. Since the predicted absolute rates are too small by one order of magnitude, \footnote{The same issue also occurs in the $D\to TP$ system \cite{CC:SAT}. The
predicted branching fractions based on factorization are at least two orders of magnitude smaller than data, even for decays free of weak annihilation contributions.  We cannot find possible sources of rate enhancement.}
the discrepancy between theory and experiment may call for the study of (i) NLO effects from vertex, hard spectator scattering and penguin corrections, (ii) power corrections from penguin annihilation, and (iii) improved estimate of $B\to T$ form factors. The detailed investigation will be left to a future work.

\section{Conclusions}

We have studied the decays $B\to K_h\eta$ and $B\to K_h\eta'$  within the framework of QCDF for $K_h= K,K^*,K_0^*(1430)$ and naive factorization for $K_h=K_2^*(1430)$. There are three different types of penguin contributions: (i) $b\to sq\bar q\to s\eta_{q}$, (ii) $b\to ss\bar s\to s\eta_s$, and (iii) $b\to s q\bar q\to q \bar K_h$, corresponding to Figs. 1(a), 1(b) and 1(c), respectively. Our conclusions are as follows:

\begin{itemize}

\item $B\to K^{(*)}\eta^{(')}$ decays are dominated by type-II and type-III penguin contributions. The interference, constructive for $K\eta'$ and $K^*\eta$ and destructive for $K\eta$ and $K^*\eta'$, between Figs. 1(b) and 1(c)  explains the pattern of $\Gamma(B\to K\eta')\gg\Gamma(B\to K\eta)$ and $\Gamma(B\to K^*\eta')\ll\Gamma(B\to K^*\eta)$. Within QCDF, the observed large rate of the $K\eta'$ mode can be naturally explained without invoking flavor-singlet contributions or something exotic. The predicted central values of the decay rates for $K^{*-}\eta'$ and $K^{*0}\eta'$ are smaller than BaBar but consistent with Belle.

\item    There are two $K^{(*)}\eta^{(')}$ modes in which direct \CP asymmetries have been measured with significance around $4\sigma$\,: $\acp(K^{-}\eta)=-0.37\pm0.09$ and $\acp(\bar K^{*0}\eta)=0.19\pm0.05$\,. In QCDF, power corrections from penguin annihilation which are needed to resolve \CP puzzles in $K^-\pi^+$ and $\pi^+\pi^-$ modes will flip $\acp(K^{-}\eta)$ into a wrong sign. We show that soft corrections to the color-suppressed tree amplitude $a_2$ in conjunction with the the charm content of the $\eta$ will finally lead to $\acp(K^-\eta)=-0.15^{+0.19}_{-0.28}$. Soft corrections to $a_2$ are also needed to improve the prediction for $\acp(\bar K^{*0}\eta)$. The QCDF prediction is in better agreement with experiment than pQCD and SCET.

\item    The decay pattern of  $B\to K_0^*(1430)\eta^{(')}$  depends on whether $K_0^*(1430)$ is an excited state of $\kappa$ or a lowest-lying $P$-wave $q\bar q$ state. If $K_0^*(1430)$ is an excited state of $q\bar q$, one will have a destructive (constructive) interference of Fig. 1(b) with Figs. 1(a) and 1(c) for $K_0^*\eta$ ($K_0^*\eta'$). This leads to $\B(B\to K_0^*\eta')\gg \B(B\to K_0^*\eta)$.
    If $K_0^*(1430)$ is made of the lowest-lying $q\bar q$, we found that Figs. 1(a) and 1(c) interfere constructively and that $A(B\to K_0^*\eta)/A(B\to K_0^*\eta')\approx \cot\!\phi$ with $\phi$ being the $\eta$-$\eta'$ mixing angle in the $\eta_q,\eta_s$ flavor basis. Hence,
    $K_0^*\eta$ has a rate slightly larger than $K_0^*\eta'$ owing to the fact that $\phi$ is less than $45^\circ$. The agreement of the latter scenario with experiment indicates that the scalar meson $K_0^*(1430)$ is indeed a bound  state of the low-lying $q\bar q$ state in $P$-wave.

\item Fig. 1(c) does not contribute to $B\to K_2^*\eta^{(')}$ under the factorization hypothesis and Fig. 1(b) dominates. The ratio $\Gamma(B\to K_2^*\eta')/\Gamma(B\to K_2^*\eta)$ is expected to be of order 2.5  as a consequence of (i) $|f_{\eta'}^s|>|f_\eta^s|$ and (ii) a destructive (constructive) interference between Figs. 1(a) and 1(b) for $K^*_2\eta$ ($K_2^*\eta'$). However, the predicted rates of $B\to K_2^*\eta^{(')}$ in naive factorization are too small by one order of magnitude and this issue remains to be resolved.

\end{itemize}

\vskip 1.71cm {\bf Acknowledgments}

One of us (H.Y.C.) wishes to thank the
hospitality of the Physics Department, Brookhaven National
Laboratory. This research was supported in part by the National
Science Council of R.O.C. under Grant Nos. NSC97-2112-M-001-004-MY3 and NSC97-2112-M-033-002-MY3.

\appendix
\section{The $\eta-\eta'$ system}
Decay constants $f^{q}_{\eta^{(')}}$, $f_{\eta^{(')}}^{s}$ and $f_{\eta^{(')}}^c$ are defined by
\be
\la 0|\bar q\gamma_\mu\gamma_5q|\eta^{(')}\ra=i{1\over\sqrt{2}}f_{\eta^{(')}}^q  p_\mu, \quad \la 0|\bar s\gamma_\mu\gamma_5s|\eta^{(')}\ra=if_{\eta^{(')}}^s p_\mu, \quad \la 0|\bar c\gamma_\mu\gamma_5c|\eta^{(')}\ra=if_{\eta^{(')}}^c p_\mu,
\en
while the widely studied decay constants
$f_q$ and $f_s$ are defined as \cite{FKS}
\begin{eqnarray}
   \langle 0|\bar q\gamma^\mu\gamma_5 q|\eta_q\rangle
   &=& \frac{i}{\sqrt2}\,f_q\,p^\mu \;, \qquad
   \langle 0|\bar s\gamma^\mu\gamma_5 s|\eta_s\rangle
   = i f_s\,p^\mu \;.\label{deffq}
\end{eqnarray}
The ansatz made by Feldmann, Kroll and Stech (FKS) \cite{FKS} is that the decay constants in the quark flavor basis follow the same pattern of $\eta-\eta'$ mixing given in Eq. (\ref{eq:qsmixing})
\begin{eqnarray}
\left(
\begin{array}{cc}
f_\eta^q & f_\eta^s \\
f_{\eta'}^q & f_{\eta'}^s
\end{array} \right)=
\left( \begin{array}{ccc}
    \cos\phi & -\sin\phi \\
    \sin\phi & \cos\phi
   \end{array} \right) \left(
\begin{array}{cc}
f_q & 0 \\
0 & f_s
\end{array} \right).
\end{eqnarray}
Empirically, this ansatz works very well \cite{FKS}. Theoretically, it has been shown recently that this assumption can be justified in the large-$N_c$ approach  \cite{Mathieu:2010ss}.

Consider the matrix elements of pseudoscalar densities \cite{BenekeETA}
\be \label{eq:hetaqs}
2m_q\la 0 |\bar q\gamma_5q|\eta^{(')}\ra={i\over\sqrt{2}}h^q_{\eta^{(')}}, \qquad
2m_s\la 0 |\bar s\gamma_5s|\eta^{(')}\ra={i}h^s_{\eta^{(')}},
\en
one can define the parameters $h_q$ and $h_s$ in analogue to $f_q$ and $f_s$
\begin{eqnarray}
   2m_q\langle 0|\bar q\gamma_5 q|\eta_q\rangle
   = \frac{i}{\sqrt2}\,h_q \;, \qquad
   2m_s \langle 0|\bar s\gamma_5 s|\eta_s\rangle
   &=& i h_s\;,\label{deffq}
\end{eqnarray}
and relate them to $h_{\eta,\eta'}^{q,s}$ by the same FKS ansatz
\begin{eqnarray}
\left(
\begin{array}{cc}
h_\eta^q & h_\eta^s \\
h_{\eta'}^q & h_{\eta'}^s
\end{array} \right)=
\left( \begin{array}{ccc}
    \cos\phi & -\sin\phi \\
    \sin\phi & \cos\phi
   \end{array} \right) \left(
\begin{array}{cc}
h_q & 0 \\
0 & h_s
\end{array} \right).
\end{eqnarray}

Using the equations of motion
\begin{eqnarray}
   \partial_\mu(\bar q\gamma^\mu\gamma_5 q) &=& 2im_q\,\bar q\gamma_5 q
   +\frac{\alpha_s}{4\pi}\,G_{\mu\nu}\,\widetilde{G}^{\mu\nu}\;,\nonumber\\
\partial_\mu(\bar s\gamma^\mu\gamma_5 s) &=& 2im_s\,\bar s\gamma_5s
   +\frac{\alpha_s}{4\pi}\,G_{\mu\nu}\,\widetilde{G}^{\mu\nu}\;,
   \label{eq:eom}
\end{eqnarray}
one can express all non-perturbative parameters in terms of the decay constants $f_q,f_s$ and the mixing angle $\phi$:
\be
h_q &=& f_q(m_\eta^2\cos^2\phi+m_{\eta'}^2\sin^2\phi)-\sqrt{2}f_s(m_{\eta'}^2-m_\eta^2)\sin\phi\cos\phi, \non \\
h_s &=& f_s(m_{\eta'}^2\cos^2\phi+m_{\eta}^2\sin^2\phi)-{f_q\over \sqrt{2}}(m_{\eta'}^2-m_\eta^2)\sin\phi\cos\phi.
\en
For numerical calculations we shall use those parameters determined from a fit to experimental data \cite{FKS}
\be
f_q=(1.07\pm0.02)f_\pi, \quad f_s=(1.34\pm0.06)f_\pi, \quad \phi=39.3^\circ\pm1.0^\circ.
\en

The masses of $\eta_q$ and $\eta_s$ read \cite{FKS}
\be
m_{\eta_q}^2 &=& {\sqrt{2}\over f_q}\la 0|m_u\bar ui\gamma_5u+m_d\bar di\gamma_5d|\eta_q\ra+{\sqrt{2}\over f_q}\la 0|{\alpha_s\over 4\pi}G\tilde G|\eta_q\ra\approx m_\pi^2+ {\sqrt{2}\over f_q}\la 0|{\alpha_s\over 4\pi}G\tilde G|\eta_q\ra , \non \\
m_{\eta_s}^2 &=& {2\over f_s}\la 0|m_s\bar si\gamma_5s|\eta_s\ra+{1\over f_s}\la 0|{\alpha_s\over 4\pi}G\tilde G|\eta_s\ra\approx 2m_K^2-m_\pi^2+ {1\over f_s}\la 0|{\alpha_s\over 4\pi}G\tilde G|\eta_s\ra,
\en
where contributions to their masses from the gluonic anomaly have been included. We shall use the parameters extracted from a phenomenological fit \cite{FKS}:
\be
&& {1\over \sqrt{2}f_q}\la 0|{\alpha_s\over 4\pi}G\tilde G|\eta_q\ra=0.265\pm0.010, \qquad
{ \la 0|{\alpha_s\over 4\pi}G\tilde G|\eta_q\ra\over \sqrt{2}\la 0|{\alpha_s\over 4\pi}G\tilde G|\eta_s\ra}={f_s\over f_q}.
\en
Numerically,
\be   \label{eq:fetaqs}
&& h_{\eta}^q=0.0013\,{\rm GeV}^3, \quad h_{\eta}^s=-0.0555\,{\rm GeV}^3, \quad h_{\eta'}^q=0.0011\,{\rm GeV}^3, \quad h_{\eta'}^s=0.068\,{\rm GeV}^3, \non \\
&& f_{\eta}^q=109\,{\rm
MeV}, \qquad~~ f_{\eta}^s=-111\,{\rm MeV}, \qquad~ f_{\eta'}^q= 89\,{\rm
MeV}, \qquad\quad~ f_{\eta'}^s=136\,{\rm MeV}, \non \\
&& f_{\eta}^c=-2.3\,{\rm MeV}, \qquad f_{\eta'}^c=-5.8\,{\rm MeV}, \qquad~~m_{\eta_q}=741\,{\rm MeV}, \qquad m_{\eta_s}=802\,{\rm MeV},
\en
where we have used the perturbative result \cite{fetac}
\be \label{eq:fetac}
f_{\eta^{(')}}^c=-{m_{\eta^{(')}}^2\over 12 m_c^2}\,{f_{\eta^{(')}}^q\over\sqrt{2}}.
\en


\end{document}